\documentclass[journal,12pt,onecolumn,draftclsnofoot,]{IEEEtran}
\usepackage{amssymb, amsthm, amscd, MnSymbol}
\DeclareMathAlphabet{\mathbbm}{U}{bbm}{m}{n}
\usepackage{graphicx}
\usepackage{nicefrac}
\usepackage{bm}

\usepackage{url}

\usepackage{cite}
\usepackage{balance}

\usepackage[para,online,flushleft]{threeparttable}
\usepackage{multirow}

\ifCLASSOPTIONcompsoc
  \usepackage[caption=false,font=normalsize,labelfont=sf,textfont=sf]{subfig}
\else
  \usepackage[caption=false,font=footnotesize]{subfig}
\fi

\usepackage{algorithmicx}
\usepackage[linesnumberedhidden,ruled,vlined]{algorithm2e}
\usepackage{color}

\usepackage{wrapfig} 
\usepackage{pgfplots}

\numberwithin{equation}{section}
\mathchardef\mhyphen="2D
\begin{document}
\title{Block Iterative Reweighted Algorithms for Super-Resolution of Spectrally Sparse Signals}
\author{{Myung~Cho, Kumar Vijay~Mishra, Jian-Feng~Cai, and Weiyu~Xu}
\thanks{Myung~Cho, Kumar Vijay~Mishra, and Weiyu~Xu are with the Department of Electrical and Computer Engineering; Jian-Feng~Cai is with the Department of Mathematics, The University of Iowa, Iowa City, IA, 52242 USA. E-mail: \{myung-cho, kumarvijay-mishra, jianfeng-cai, weiyu-xu\}@uiowa.edu.}}
\maketitle
\begin{abstract}
We propose novel algorithms that enhance the performance of recovering unknown continuous-valued frequencies from undersampled signals.  Our iterative reweighted frequency recovery algorithms employ the support knowledge gained from earlier steps of our algorithms as block prior information to enhance frequency recovery.  Our methods improve the performance of the atomic norm minimization which is a useful heuristic in recovering continuous-valued frequency contents. Numerical results demonstrate that our block iterative reweighted methods provide both better recovery performance and faster speed than other known methods.
\end{abstract}
\IEEEpeerreviewmaketitle

\section{Introduction}
\label{sec:intro}
Compressed sensing promises to perform signal recovery using a smaller number of samples than required by the Nyquist-Shannon sampling theorem. In the compressed sensing framework, a sparse signal $x$ is recovered from the observation vector $y$ even though the dimension of $y$ is much smaller than the dimension of $x$. Since compressed sensing reduces the sampling rate in recovering sparse signals, it has made great impacts in various signal processing areas \cite{eldar2012compressed}.

Compressed sensing has also found application to the problem of \textit{line spectral estimation}, which aims to estimate spectral information from few observations. Early-stage compressed sensing frameworks for spectral estimation \cite{duarte2013spectral, chi2011sensitivity} assumed that the frequencies of spectrally sparse signals were located on discretized grid points in the frequency domain. However, in practice, frequencies can take values in a continuous domain, giving rise to the so-called \textit{basis mismatch} problem \cite{chi2011sensitivity} when the discretization of the frequency domain is not fine enough.

The breakthrough theory of \textit{super-resolution} \cite{candes2014towards} proposed by Cand\`{e}s and Fernandez-Granda states that sparse continuous-valued frequencies can be exactly recovered through total variation minimization using a set of $n$ \textit{uniformly} spaced time samples, provided the minimum separation between any two frequencies is $4/n$. In order to recover continuous-valued frequencies from few \textit{randomly} chosen nonuniformly-spaced time samples, Tang \textit{et al.} proposed off-the-grid compressed sensing that employs atomic norm minimization for frequency recovery \cite{tang2012csotg}. Later, it was shown that the $\ell_1$ minimization over the fine discrete dictionary provides an approximate solution to the atomic norm minimization \cite{tang2013justdiscretize}.

In this paper, we are interested in recovering spectrally sparse signals with as few random time samples as possible. It is then natural to ask whether there are efficient frequency recovery algorithms that can further improve the performance or relax the frequency separation conditions when compared with the total variation minimization or atomic norm minimization. We propose new iterative algorithms to enhance the performance of recovering continuous-valued frequency. In our iterative algorithms, we estimate the frequency support information from previous iterations, and use the support information as \textit{block prior} \cite{mishra2014Super} for reweighted atomic norm minimization in later iterations. Numerical results show that we can improve recovery performance by exploiting the block prior provided in earlier iterations.

We remark that there are quite a few works in the literature\cite{chartrand2008iteratively,candes2008enhancing,wipf2010iterative,needell2009noisy,fang2014super,yang2014enhancing} where iterative reweighted methods have been used to improve sparse recovery performance in compressed sensing. However, the sparse signal recovery is considered over a finite discrete dictionary in \cite{chartrand2008iteratively,candes2008enhancing,wipf2010iterative,needell2009noisy,fang2014super}. Besides \cite{yang2014enhancing}, only our work considers recovering continuous-valued frequencies by directly reweighting in the continuous dictionary through a semi-definite program (SDP). Our work differs from \cite{yang2014enhancing} in that we provide different reweighting schemes that lead to improved signal recovery performance. In \cite{yang2014enhancing}, the authors set the reweighting weight $w(f)$ for a frequency $f \in [0,1]$ according to correlations between frequency atoms (see e.g. Theorem 3 of \cite{yang2014enhancing}).  In contrast, our method allows $w(f)$ to take more general forms through the dual program of weighted atomic minimization under general weights  \cite{mishra2014spectral}, thereby lending more flexibility to incorporating external prior information  and prior information passed on from earlier algorithm iterations. Numerical experiments show that our iterative algorithms improve both the recovery performance and the execution time, compared with \cite{tang2012csotg} and \cite{yang2014enhancing}.

\section{Background on Standard and Weighted Atomic Norm Minimization Algorithms}
\label{sec:BITERANM}
In this paper, we denote the set of complex numbers, real numbers, positive integers and natural numbers including $0$ as $\mathbb{C}$, $\mathbb{R}$, $\mathbb{Z}^{+}$, and $\mathbb{N}$ respectively. We reserve calligraphic uppercase letters for index sets. When we use an index set $\mathcal{K}$ as the subscript of a vector $x$ or a matrix $F$, i.e., $x_{\mathcal{K}}$ or $F_{\mathcal{K}}$, it represents the part of the vector $x$ over index set $\mathcal{K}$ or the columns of the matrix $F$ over index set $\mathcal{K}$ respectively.

Let $x^{\star}$ be a spectrally sparse signal expressed as a sum of $k$ complex exponentials as follows:
\par\noindent
\small
\begin{align}
\label{eq:sigmodelstd}
    x^{\star}_l = \sum_{j=1}^{k} c^{\star}_j e^{i2\pi f^{\star}_j l} = \sum_{j=1}^{k} |c^{\star}_j|a(f^{\star}_j, \phi^{\star}_j)_l,\;\;\;\;  l \in \mathcal{N},\\[-20pt] \nonumber
\end{align}
\normalsize
where $f^{\star}_j \in [0,1]$ represents a frequency, $c^{\star}_j = |c^{\star}_j|e^{i\phi^{\star}_j}$ is its coefficient,  and $\phi^{\star}_j \in [0,2\pi]$ is its phase, $\mathcal{N} = \{l\; :\; 0 \leq l \leq n-1, l \in \mathbb{N} \}$ is the set of time indices. Here, $a(f^{\star}_j, \phi^{\star}_j) \in \mathbb{C}^{|\mathcal{N}|}$ is a \textit{frequency-atom}, with the $l$-th element given by $a(f^{\star}_j, \phi^{\star}_j)_l = e^{i(2\pi f^{\star}_j l + \phi^{\star}_j)}$. In particular, when phase is $0$, we denote the frequency-atom simply as $a(f_j)$. We assume that the signal in (\ref{eq:sigmodelstd}) is observed over the time index set $\mathcal{M} \subseteq \mathcal{N}$, $|\mathcal{M}| = m \leq n$, where $m$ observations are chosen randomly. Our goal is to recover all the frequencies with the smallest possible number of observations. Estimating frequencies is not trivial because they are in continuous domain, and their phases and magnitudes are also unknown.

The atomic norm of a signal $x$ and its dual norm \cite[Eq. (II.7)]{tang2012csotg} are defined respectively as follows:
\par\noindent
\small
\begin{align}
\label{eq:atomicNorm}
& ||x||_{\mathcal{A}} = \inf \{ \sum_j |c_j|: x = \sum_j c_j a(f_j)\},\\[-5pt]
\label{eq:daulAtomicNorm}
& ||q||^{*}_{\mathcal{A}} = \sup_{||x||_{\mathcal{A}} \leq 1 } \langle q, x \rangle_{\mathbb{R}}  =  \sup_{\substack{\phi \in [0,2\pi],\\ f \in [0,1]}} \langle q, e^{i\phi}a(f,0) \rangle_{\mathbb{R}} = \sup_{f \in [0,1]} |\langle q, a(f) \rangle|,\nonumber \\[-15pt]
\end{align}
\normalsize
where $\langle q, x \rangle_{\mathbb{R}}$ represents the real part of the inner product $x^H q$. Here, the superscript $H$ is used for the conjugate transpose. In \cite{tang2012csotg}, the authors proposed the following atomic norm minimization to recover a spectrally sparse signal $x^{\star}$ using randomly chosen time samples $\mathcal{M} \subseteq \mathcal{N}$:
\par\noindent
\small
\begin{align}
\label{eq:atomicNormMin}
	\underset{x}{\text{minimize}}\;\;  ||x||_{\mathcal{A}} \;\;\;\; \text{subject to}\;\; x_j = x^{\star}_j, \;\; j \in \mathcal{M}. \\[-20pt] \nonumber
\end{align}
\normalsize
The dual problem of (\ref{eq:atomicNormMin}) is
\par\noindent
\small
\begin{align}
	\label{eq:atomicNormMinDaul}
	 \underset{q}{\text{maximize}} & \;\; \langle q_{\mathcal{M}},x^{\star}_{\mathcal{M}}\rangle_{\mathbb{R}} \;\;\;\; \text{subject to}\;\; q_{\mathcal{N}\setminus \mathcal{M}}=0,\; ||q||^{*}_{\mathcal{A}} \leq 1.\\[-20pt] \nonumber
\end{align}
\normalsize
The constraint $||q||^{*}_{\mathcal{A}} \leq 1$ in (\ref{eq:atomicNormMinDaul}) can be changed to $\sup_{f \in [0,1]} | \langle q, a(f)\rangle| \leq 1$ using (\ref{eq:daulAtomicNorm}). We label $\langle q, a(f) \rangle$ as the \textit{dual polynomial} $Q(f)$. Since the Slater condition is satisfied in (\ref{eq:atomicNormMin}), there is no duality gap between (\ref{eq:atomicNormMin}) and (\ref{eq:atomicNormMinDaul})   \cite{boyd2004convex}. Moreover, the estimated spectral content comprises the frequencies at which the absolute value of the dual polynomial, which is derived from $q$ (obtained as a solution of (\ref{eq:atomicNormMinDaul})), attains the maximum modulus of unity. We refer the reader to \cite{tang2012csotg} for details. The off-the-grid compressed sensing approach in \cite{tang2012csotg} demonstrated that with randomly chosen observation data, one can correctly obtain frequency information by solving the atomic norm minimization. However, the atomic norm minimization requires a certain minimum separation between frequencies for successful recovery.

In \cite{mishra2014Super}, we considered frequency recovery with external prior information, and showed that if the frequencies are known to lie in frequency subbands, we can obtain better recovery performance by using frequency block prior information. The SDP formulation adopted inside each iteration of our new algorithms follows that detailed in \cite{mishra2014Super}. We summarize that SDP formulation in the following paragraph.

Suppose the frequency $f$ of the signal $x$ lies within the frequency block $\mathcal{B} \subset [0,1]$. Then, given this block prior information $\mathcal{B}$, the atomic norm with block priors and its dual are stated respectively as follows \cite[Eq. (II.9)]{mishra2014Super}:
\par\noindent
\small
\begin{align}
\label{eq:blkWeightedAtomicNorm}
    & ||x||_{\mathcal{A,B}} = \inf \{ \sum_j |c_j|: x = \sum_j c_j a(f_j),\;f_j \in \mathcal{B}\},\\[-5pt]
\label{eq:blkDaulWeightedAtomicNorm}
    & ||q||^{*}_{\mathcal{A,B}} =  \sup_{||x||_{\mathcal{A,B}}\leq 1} \langle q, x \rangle_{\mathbb{R}} =  \sup_{f \in \mathcal{B}} |\langle q, a(f) \rangle|. \\[-20pt] \nonumber
\end{align}
\normalsize
We formulate the atomic norm minimization with block priors \cite[Eq. (III.1)]{mishra2014Super} as
\par\noindent
\small
\begin{align}
	\label{eq:blockpriorprimal}
	& \underset{x}{\text{minimize}}\;\;  ||x||_{\mathcal{A}, \mathcal{B}}\;\;\;\; \text{subject to}\;\; x_j = x_j^{\star}, \;\; j \in \mathcal{M}.
\end{align}
\normalsize
The dual problem of (\ref{eq:blockpriorprimal}) is
\par\noindent
\small
\begin{align}
	\label{eq:dual_substituted}
	 \underset{q}{\text{maximize}} \; \langle q_{\mathcal{M}},x_{\mathcal{M}}\rangle_{\mathbb{R}}\;\;\text{subject to}\;\; q_{\mathcal{N}\setminus \mathcal{M}}=0,\; ||q||^{*}_{\mathcal{A,B}} \leq 1.
\end{align}
\normalsize
Here, $\mathcal{B}$ is a union of disjoint frequency blocks within which all the true frequencies are located, i.e., $\displaystyle f^{\star}_j \in \mathcal{B}, \;\;\mathcal{B} = \cup^r_{k=1} [f_{L_k}, f_{H_k}]$, where $r$ is the number of disjoint block blocks, $f_{L_k}$ and $f_{H_k}$ are the lowest and highest frequencies of the $k$-th frequency block. Using the properties of positive trigonometric polynomials \cite{fejer1915uber,dumitrescu2007positive} and (\ref{eq:blkDaulWeightedAtomicNorm}), this dual problem can be formulated as an SDP \cite[Eq. (III.16)]{mishra2014Super}:
\par\noindent
\small
\begin{align}
	\label{eq:blocksparsitySDP}
	& \underset{\begin{subarray}{c}
	 q,\{\bm{G}_{a_i}\}_{i=1}^r, \{\bm{G}_{b_i}\}_{i=1}^r\\
	\end{subarray}}{\text{maximize}} \; \langle q_{\mathcal{M}},x_{\mathcal{M}} \rangle_{\mathbb{R}} \nonumber\\
	& \;\;\; \text{subject to}  \;\; q_{\mathcal{N}\setminus \mathcal{M}}=0,  \\
	& \quad\quad\quad\quad \;\;\;\; \delta_{k_i} = \mathcal{L}_{k_i, f_{L_i}, f_{H_i}}(\bm{G}_{a_i}, \bm{G}_{b_i}),\; \substack{{k_i=0,...,(n-1),}\\{i=1,...,r}},\nonumber \\
    & \quad\quad\quad\quad \;\;\;\; \begin{bmatrix} \bm{G}_{a_i} & q \\ q^{H} & 1 \end{bmatrix} \succeq 0,\;\; i=1,...,r, \nonumber
\end{align}
\normalsize
where $\delta_{k_i} = 1$ if $k_i=0$, and $\delta_{k_i} = 0$ otherwise, $\bm{G}_{a_i} \in \mathbb{C}^{n \times n}$ and $\bm{G}_{b_i} \in \mathbb{C}^{(n-1) \times (n-1)}$ are Gram matrices. The trace parameterization term $\mathcal{L}_{k, f_{L}, f_{H}}(\bm{G}_{a}, \bm{G}_{b})$ for the frequency block $[f_L, f_H] \subset [0, 1]$ is set to $\text{tr}{[\mathbf{\Theta}_k \bm{G}_{a}]} + \text{tr}{[(d_1\mathbf{\Theta}_{k-1} + d_0\mathbf{\Theta}_{k} + d_1^{H}\mathbf{\Theta}_{k+1}) \cdot \bm{G}_{b}]}$, where $\mathbf{\Theta}_k$ is the Toeplitz matrix that has ones on the $k$-th diagonal and zeros elsewhere, $d_0 = -\frac{\alpha \beta + 1}{2}$, $d_1 = \frac{1-\alpha \beta}{4} + i \frac{\alpha + \beta}{4}$, where $\alpha = \tan(2\pi f_L /2)$, and $\beta = \tan(2\pi f_H/2)$ when $[f_L, f_H] \subset [0, 0.5]$, and $\alpha = \tan(2\pi (f_L-1) /2)$, and $\beta = \tan(2\pi (f_H-1)/2)$ when $[f_L, f_H] \subset (0.5, 1]$. This SDP approach \cite{mishra2014Super} was expanded for more general cases in \cite{mishra2014spectral}.


Although the atomic norm minimization that exploits external prior information can improve signal recovery performance, in practice, one may not always have direct access to prior information. This leads to the question if we can improve the frequency recovery performance without any external prior information. We describe new algorithms to address this issue in the following section.

\section{Block iterative (re)weighted Atomic Norm Minimization Algorithms}
\label{sec:BITMix}
We propose three iterative algorithms to enhance frequency recovery performance in the absence of external prior information. In our algorithms, we use estimated frequency support information from previous iterations as block prior for subsequent iterations.

\subsection{Block iterative weighted Atomic Norm Minimization}
We first introduce a conceptual algorithm named Block iterative weighted Atomic Norm Minimization (BANM). BANM solves SDP of (\ref{eq:blocksparsitySDP}) repeatedly, using block priors obtained from the previous iteration. In each iteration, BANM estimates the frequency locations,  and then, around the estimated frequencies, BANM forms blocks which very likely contain the true frequencies.  With the block priors so obtained, BANM enhances frequency recovery via solving (\ref{eq:blocksparsitySDP}) in the next iteration, using the new block information.

BANM initially sets the iteration number $t = 0$, frequency block $\mathcal{B} = [0,1]$, $f \in \mathcal{B}$, and then solves (\ref{eq:blocksparsitySDP}). Suppose the solution of (\ref{eq:blocksparsitySDP}) gives $r$ estimated frequencies $f_i^{(t)}$, $i=1,...,r$, where the superscript $(t)$ is used to represent the iteration number. BANM chooses the $l$ frequencies $f_{i_1}$, $f_{i_2}$, ...,  $f_{i_l}$ with the largest coefficients in amplitude among them, where $l$ is a certain integer number. BANM then forms a union frequency block $\mathcal{B}$ with $l$ frequency subbands around the estimated frequencies $f_{i_j}^{(t)}$, $j=1,...,l$, as
\vspace*{-0.20cm}
\par\noindent
\small
\begin{align}
\label{eq:freq_band}
   \mathcal{B} = \bigcup^{l}_{j=1} [f_{i_j}^{(t)}-\tau, f_{i_j}^{(t)}+\tau],\\[-20pt] \nonumber
\end{align}
\normalsize
for some small real number $\tau > 0$, $\tau \in [0,1]$ that determines the size of the subband. BANM uses the union frequency block $\mathcal{B}$ as block prior and solves (\ref{eq:blocksparsitySDP}) again with updated parameters. The algorithm continues solving (\ref{eq:blocksparsitySDP}) and updating (\ref{eq:freq_band}) in each iteration until either a maximum number of iterations is reached or the solution of (\ref{eq:blocksparsitySDP}) converges.

\subsection{Block iterative reweighted $\ell_1$ and Atomic Norm Minimization Mixture}
BANM requires solving SDP in each iteration to estimate the location of frequencies, but solving SDP repeatedly causes long execution time. Thus, we propose using low-complexity algorithm to obtain prior information on frequency locations. We then use the aforementioned SDP (\ref{eq:blocksparsitySDP}) only in the last iteration for accurately determining the frequency locations. This concept is the key to design of our algorithm - Block iterative reweighted $\ell_1$ and Atomic Norm Minimization Mixture (or simply, BANM-Mix) algorithm that can achieve super-resolution of frequencies with low complexity (Algorithm \ref{alg:BITERL1}).

BANM-Mix first discretizes the continuous frequency domain [0,1] in uniform intervals of size $\bigtriangleup_f$. We denote the index set for these intervals as $\mathcal{P}=\{i\}_{i=1}^p$, where $p = 1/\bigtriangleup_f$. The index set corresponds to $p$ discrete frequency grid points $f_j=(j-1)/p$, $1\leq j\leq p$. We have the discrete Fourier matrix $F \in \mathbb{C}^{n \times p}$ over $p$ discrete frequency grid points whose element in the $j$-th column and $l$-th row is $a(f_j)_l = e^{i(2\pi f_j l)}$.

Then, BANM-Mix iteratively solves reweighted $\ell_1$ minimization over this discretized frequency dictionary to efficiently estimate frequency locations. Different from iterative reweighted $\ell_1$ minimization algorithms designed for incoherent discrete dictionaries \cite{chartrand2008iteratively,candes2008enhancing,wipf2010iterative,needell2009noisy,fang2014super}, our iterative reweighted $\ell_1$ minimization algorithm employs novel \emph{adaptive gridding} and \emph{block reweighting} strategies to extract frequency support information from our highly correlated discretized dictionary.

BANM-Mix initializes coefficients $c^{(0)}_i = 0$, weights $w_i ^{(0)}= 1$ for $i=1,...,p$, and an index set $\mathcal{K}^{(0)} = \{j: j=ql+1,\; l=0,1,...,(p-1)/q,\; q \in \mathbb{Z}^{+} \} \triangleq \{j\}_{j=1:q:p} \subseteq \mathcal{P}$. Let $W^{(t)} = diag(w^{(t)}_{\mathcal{K}^{(t)}})$ be a diagonal matrix with weights $w_{\mathcal{K}^{(t)}}^{(t)}$, $F_{\mathcal{K}^{(t)}} \in \mathbb{C}^{|\mathcal{M}| \times |\mathcal{K}^{(t)}|}$ be the partial discrete Fourier matrix.

In the $t$-th iteration, BANM-Mix solves the following weighted $\ell_1$ minimization problem over the index set $\mathcal{K}^{(t)}$, rather than the larger index set $\mathcal{P}$:
\par\noindent
\small
\begin{align}
        \label{eq:weightedL1}
            & \underset{z}{\text{minimize}}\;\; || W^{(t)} z ||_1 \quad \text{subject to} \;\; x^{\star}_l = (F_{\mathcal{K}^{(t)}} z)_l,\;\;l \in \mathcal{M}.           \\[-20pt] \nonumber
\end{align}
\normalsize
We then define a vector $c^{(t)}$ having $c^{(t)}_{\mathcal{K}^{(t)}}=z$ and $c^{(t)}_{\mathcal{P}\setminus {\mathcal{K}^{(t)}}}=0$.
\begin{figure}[t]
    \centering
     \includegraphics[scale=0.5]{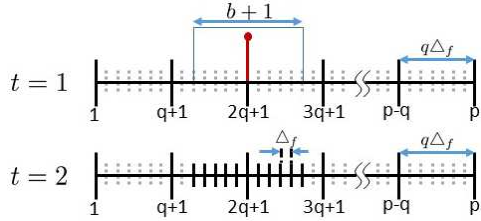} 
    \caption{An illustration of the adaptive gridding. The estimated frequency $f_{2q+1}$ in the first iteration is depicted by a red pole. The index set $\mathcal{K}$ and $\mathcal{P}$ have solid and dotted grid points respectively. The block that contains the red pole in the middle is the index-wise frequency block $\mathcal{BI}_{2q+1}$, where $b=10$.}
    \label{fig:adaptive_grid}
\end{figure}

BANM-Mix then calculates the weight $w_i^{(t+1)}$, $i=1,...,p$. We define the index-wise frequency block $\mathcal{BI}_i$ as\\[-15pt]
\par\noindent
\small
\begin{align}
\label{eq:rangeOfFreq}
   & \mathcal{BI}_i = \{j\; :\; i-b/2 \leq j \leq  i+b/2,\; j\in \mathcal{P} \},\\[-15pt] \nonumber
\end{align}
\normalsize
for some positive integer $b$, which determines the block width. BANM-Mix computes the weight $w_i^{(t+1)}$ by considering the frequency coefficients around $f_i$ in the discretized domain as
\vspace*{-0.10cm}
\par\noindent
\small
\begin{align}
\label{eq:updateWeight}
\displaystyle
    w_i^{(t+1)} = \frac{1}{\sum_{j \in \mathcal{BI}_i} | c_j^{(t)} | + \epsilon}, \;\; i = 1,...,p \\[-20pt] \nonumber
\end{align}
\normalsize
where $\epsilon$ is a small positive constant to prevent $w_i^{(t+1)}$ from going to infinity. We refer to our procedure in (\ref{eq:updateWeight}) as \textit{block reweighting}. Since the discretized dictionary under consideration has highly correlated columns, block reweighting can accurately reflect the likelihood of a true frequency existing around $f_{i}$. Our numerical experiments showed that earlier reweighting strategies \cite{chartrand2008iteratively,candes2008enhancing,wipf2010iterative,needell2009noisy}, which update $w_i \leftarrow \frac{1}{|c_i| + \epsilon}$, could not correctly reflect the likelihood of a true frequency being at index $i$ and resulted in worse frequency recovery performance. This is because the solution to (\ref{eq:weightedL1}) will disperse the amplitude of a true frequency into the neighboring indices in highly correlated dictionary columns.

After updating $w_i^{(t+1)}$ for $i=1,...,p$, BANM-Mix updates the index set $\mathcal{K}^{(t+1)}$ through adaptive gridding. In adaptive gridding, BANM-Mix first finds indices $i$, $1\leq i \leq p$, with $w_i^{(t+1)} < \nicefrac{(\min(w^{(t+1)})+\max(w^{(t+1)}))}{2}$, where $\min(w^{(t+1)})$ and $\max(w^{(t+1)})$ are the minimum and maximum values of the elements of $w^{(t+1)}$ respectively. We define $\nicefrac{(\min(w^{(t+1)})+\max(w^{(t+1)}))}{2}$ as $w^{(t+1)}_{\text{mid}}$. Then BANM-Mix updates $\mathcal{K}^{(t+1)}$ as
\vspace*{-0.10cm}
\par\noindent
\small
\begin{align}
\label{eq:update_indexset}
    \mathcal{K}^{(t+1)} = \mathcal{K}^{(t)}\bigcup \bigg( \bigcup_{\{i:\; w_i^{(t+1)} < w^{(t+1)}_{\text{mid}},\; i\in \mathcal{P}\}} \mathcal{BI}_i \bigg). \nonumber \\[-16pt]
\end{align}
\normalsize
Namely, if $w_i^{(t+1)} < w^{(t+1)}_{\text{mid}}$, $\mathcal{K}^{(t+1)}$ will include finer grid points (with separation $\bigtriangleup_f$) around frequency $(i-1)/p$. Recall that, at the beginning, $\mathcal{K}^{(0)}$ has only grid points with separation $q \bigtriangleup_f$.  The reason is that when $w_i^{(t+1)}$ is small, very likely a true frequency exists around frequency $(i-1)/p$. By applying finer gridding around frequency $(i-1)/p$, one can estimate the frequency location more accurately in the next iteration. We call this method of applying different resolutions in the discretized dictionary as  \textit{adaptive gridding} (see Fig. \ref{fig:adaptive_grid}).
\begin{algorithm}[t]
\LinesNumbered
  \caption{Block iterative reweighted $\ell_1$ and Atomic Norm Minimization Mixture (BANM-Mix) Algorithm}
  \label{alg:BITERL1}
  \SetAlgoLined
{\footnotesize
   \KwIn{ $F \in \mathbb{C}^{n\times p}$, $x^{\star}_{\mathcal{M}}$, \textit{MaxItr}, $b$, $\epsilon$, $\epsilon_{err}$}
   \KwOut{ frequency $\hat{f}$, coefficient $\hat{c}$}
   \textbf{Initialize:} $t \leftarrow 0$, $c^{(t)} \leftarrow 0$, $w^{(t)} \leftarrow 1$, $\mathcal{K}^{(t)} \leftarrow \{i\}_{i=1:q:p}$, $\mathcal{P} \leftarrow \{i\}_{i=1}^p$ \par
   \For { $t=1$  \KwTo MaxItr }
   {
       $c^{(t)}_{\mathcal{K}^{(t)}}$  $\leftarrow$ solution of (\ref{eq:weightedL1}), $c^{(t)}_{\mathcal{P} \setminus \mathcal{K}^{(t)}}$ $\leftarrow$ $0$ \par
       $\mathcal{BI}_i$ $\leftarrow$ frequency block via (\ref{eq:rangeOfFreq}) for $i=1,...,p$ \par
       $w^{(t+1)}_{i}$ $\leftarrow$ weight via (\ref{eq:updateWeight}) for $i=1,...,p$ \par
       \If {$ || c^{(t-1)} - c^{(t)} ||_2$ $<$ $\epsilon_{err}$ }
       {
            $\mathcal{B}$ $\leftarrow$ frequency block via (\ref{eq:freq_band}), where  $f_i$ satisfying $w^{(t+1)}_i < \nicefrac{(\min(w^{(t+1)})+\max(w^{(t+1)}))}{2}$  \par
            $\hat{f}$ $\leftarrow$ $f$ such that $|Q(f)| = 1$ in $\mathcal{B}$ after solving (\ref{eq:blocksparsitySDP}) \par
            $\hat{c}$ $\leftarrow$ $c$ satisfying linear equation (\ref{eq:sigmodelstd}) with given $\hat{f}$ and $x^{\star}_{\mathcal{M}}$\par
            break \par
       }
       $\mathcal{K}^{(t+1)}$ $\leftarrow$ index set via (\ref{eq:update_indexset})\par
   }
}%
\end{algorithm}

The algorithm continues solving (\ref{eq:weightedL1}) in each iteration until either a specified maximum number of iterations (MaxItr) is exhausted or the solution of (\ref{eq:weightedL1}) converges i.e., $|| c^{(t-1)} - c^{(t)} ||_2 \leq \epsilon_{err}$, for some error tolerance $\epsilon_{err} > 0$. BANM-Mix then chooses the block prior set $\mathcal{B}$ by a union of the frequency blocks around frequency $f_i^{(t)}$ satisfying $w_i^{(t+1)}  < w_{\text{mid}}^{(t+1)}$. With this frequency block information, we use SDP (\ref{eq:blocksparsitySDP}) to super-resolve frequencies in the last iteration.

\subsection{Block iterative reweighted $\ell_1$ Minimization}
The complexity of BANM-Mix can still be high since we have to solve an SDP in the last iteration. To further reduce its
complexity, we propose the Block iterative reweighted $\ell_1$ Minimization (BL1M) algorithm which is the same as BANM-Mix except that BL1M does not solve SDP in the last iteration. Instead, BL1M uses postprocessing to estimate the final frequencies from the results of iterative reweighted $\ell_1$ minimizations. In the last iteration, BL1M finds the frequency blocks $\mathcal{BI}_i$ that satisfy $w_i^{(t+1)} < w_{\text{mid}}^{(t+1)}$. If two frequency blocks $\mathcal{BI}_i$ overlap, BL1M merges them into one. BL1M assumes that one frequency block contains only one true frequency. Suppose that one frequency block (after possible merging) has $r$ grid frequencies $f_1,..., f_r$ whose corresponding coefficients are $c_1,..., c_r$. Then BL1M estimates the frequency $\hat{f}$ in that block as $\hat{f} = \frac{\sum_{i=1}^{r} f_i\times |c_i|}{\sum_{i=1}^{r}|c_i|}$.

\section{Numerical Experiments}
\label{sec:experiment}
We compare our algorithms with the standard Atomic Norm Minimization (ANM) \cite{tang2012csotg}, and the Reweighted Atomic norm Minimization (RAM) \cite{yang2014enhancing}. We use CVX \cite{cvx} to solve convex programs.\footnote{We conducted our numerical experiments on HP Z220 CMT with Intel Core i7-3770 dual core CPU @3.4GHz clock speed and 16GB DDR3 RAM, using Matlab (R2013b) on Windows 7 OS.} In
all experiments, the phases and frequencies are sampled uniformly at random in $[0,2\pi)$ and $[0,1]$ respectively. The amplitudes $|c_j|$, $j = 1,...,k$, are drawn randomly from the distribution $\sqrt{0.5 + \chi^2_1}$ where $\chi^2_1$ represents the chi-squared distribution with 1 degree of freedom.

\begin{figure}[!t]
\centering
  \includegraphics[width=4.0 in]{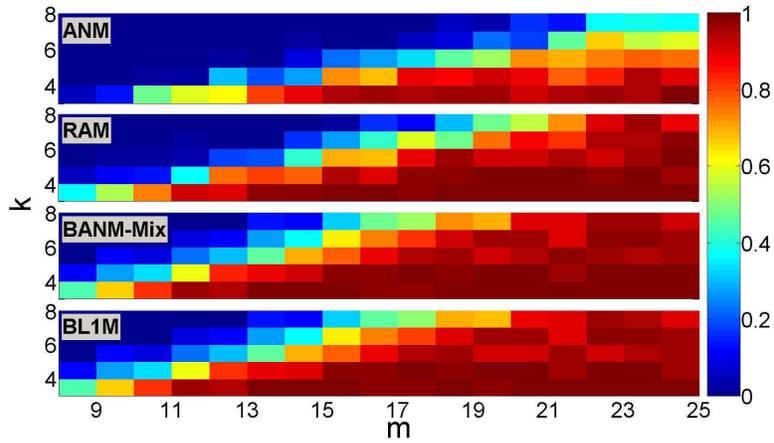}
  \caption{\small The probability $P$ of successful frequency recovery ($n = 64$).}
\label{fig:recoveryPerformance1}
\end{figure}

\begin{figure}[!t]
\centering
  \includegraphics[width=2.9 in]{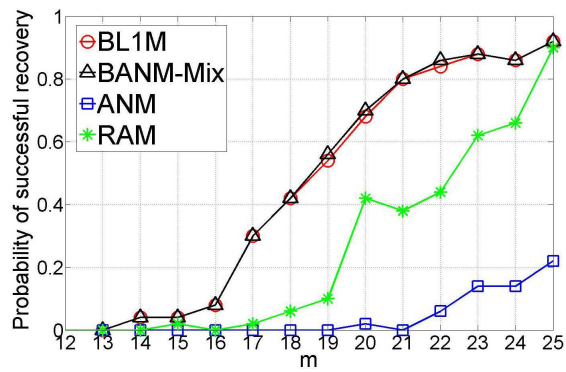}
  \caption{\small The probability $P$ of frequency recovery for $(n,k) = (64,8)$.}
\label{fig:recoveryPerformance2}
\end{figure}

%
We evaluate the recovery performance for the signal dimension $n = 64$, number of observation $m$ is varied from $8$ to $25$, block width $b = 20$, $\epsilon = 2^8$, and $\epsilon_{err} = 0.5 \times 10^{-4}$. The maximum number of iterations (MaxItr) is set to 20 for both BANM-Mix and RAM.\footnote{A MaxItr value of 20 was sufficient to guarantee an empirical convergence of our iterative procedures in most of our experiments.} Fig. \ref{fig:recoveryPerformance1} and \ref{fig:recoveryPerformance2} show the probability of successful recovery of the entire spectral content over $50$ trials for each parameter setup. We consider a recovery successful if $||f^{\star} - \hat{f}||_2 \leq 10^{-3}$. Fig. \ref{fig:recoveryPerformance2} clearly shows that our algorithm outperforms both ANM and RAM for $n=64$ and $k=8$.

\begin{figure}[!t]
\centering
  \includegraphics[scale=0.152]{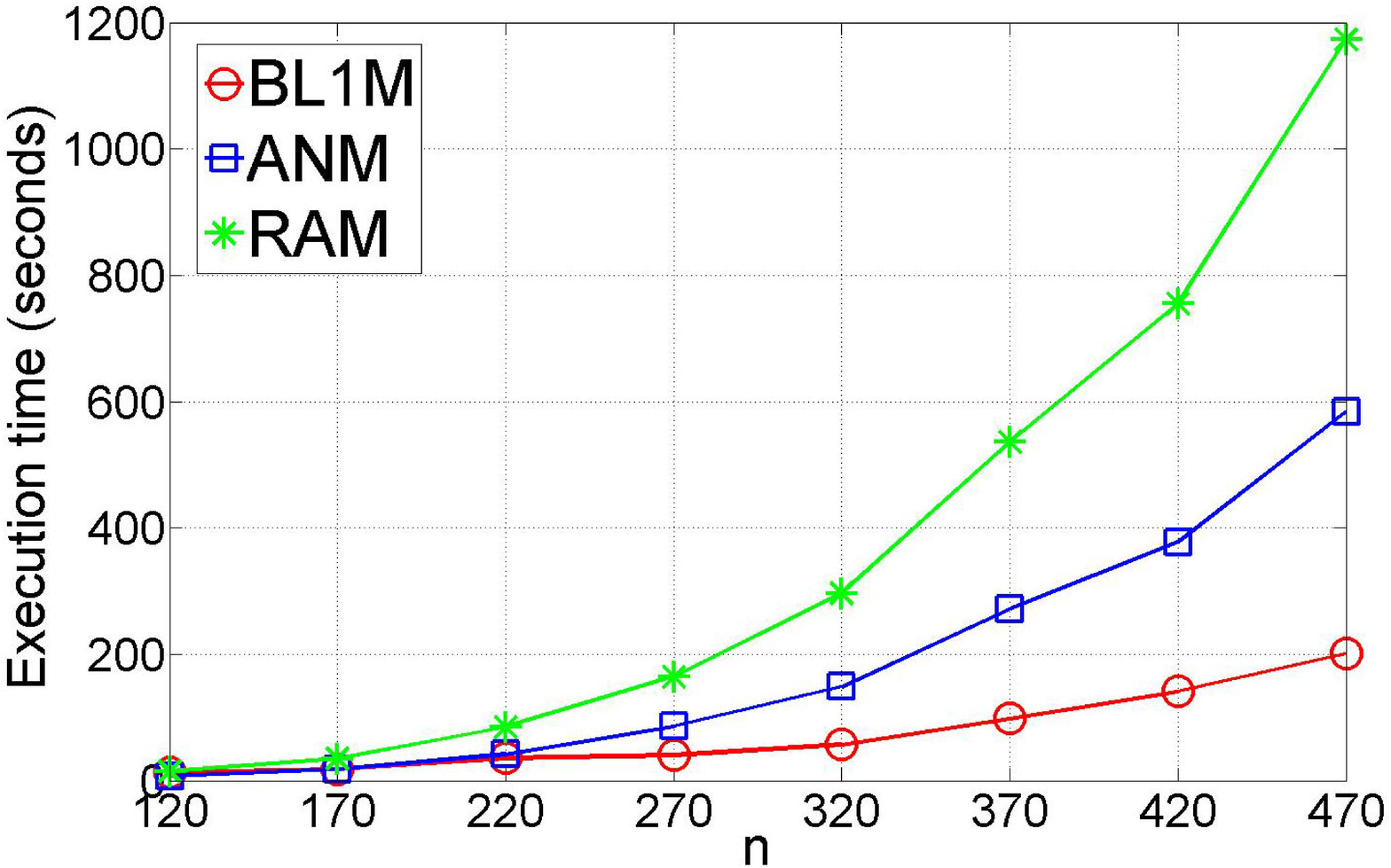}
  \caption{\small The execution time as a function of signal dimension $n$.}
\label{fig:complexity}
\end{figure}

We assess the computational complexity of algorithms in terms of the average execution time for signal recovery from 10 trials. Here, we present results when $n$ is from 120 to 470,  $m = \lfloor \nicefrac{n}{2} \rfloor$, $q=2^4$, $p = 2^{14}$, $b = 20$, $\epsilon = 2^8$, $\epsilon_{err} = 0.5 \times 10^{-4}$. Fig. \ref{fig:complexity} shows that the speed of BL1M is faster than that of ANM and RAM. This is because the latter is based on an SDP while the former uses only $\ell_1$ minimization.
\section{Conclusion}
The BANM-Mix and BL1M show better recovery than other known iterative methods \cite{yang2014enhancing,tang2012csotg}. In particular, BL1M has shorter execution times than these other methods. Our simulations empirically exhibit convergence of our iterative procedures. It would be interesting to perform more comprehensive theoretical analysis of convergence in the future.
\section*{Acknowledgement}
We thank Yuejie Chi of Ohio State University and Zai Yang of Nanyang Technological University for helpful discussions.

\bibliographystyle{IEEEbib}

\end{document}